\documentclass{elsart}

 \usepackage{graphicx}

\usepackage{amssymb}

\newcommand{\ETAL}{{\it et al.}}

\newcommand{\QUINT}{{\rm \scriptscriptstyle Q}}

\newcommand{\CONST}{{\rm Const}}

\newcommand{\BAR}{{\rm b}}
\newcommand{\MAT}{{\rm mat}}

\def\gtrsim{\;\raise 0.4ex\hbox{$>$}\kern -0.7em\lower 0.62
ex\hbox{$\sim$}\;}
\def\lesssim{\;\raise 0.4ex\hbox{$<$}\kern -0.8em\lower 0.62
ex\hbox{$\sim$}\;}

\newcommand{\Obhh}{\Omega_\BAR h^2}
\newcommand{\sig}{\sigma_8}
\newcommand{\OQ}{\Omega_\QUINT}
\newcommand{\OmM}{\Omega_\MAT}

\newcommand{\OmL}{\Omega_{\lambda}}
\newcommand{\OmT}{\Omega_{\rm tot}}
\begin{document}

\begin{frontmatter}



\title{Cosmological constraints in $\Lambda$-CDM and Quintessence paradigms with Archeops}


\author[lab1]{Marian Douspis\thanksref{label2}}
\ead{douspis@astro.ox.ac.uk}
\author[lab2]{Alain Riazuelo}\author[lab3,lab4]{, Yves Zolnierowski}\author[lab3]{, Alain Blanchard}

\& \author{the Archeops collaboration}
\ead[url]{http://www.archeops.org/}
\address[lab1]{Astrophysics, Keble Road, OX1 3RH Oxford (UK)}
\address[lab2]{CEA/DSM/SPhT, CEA/Saclay, F--91191 Gif-sur-Yvette c\'edex
(France) }
\address[lab3]{LAOMP, 14 Avenue E.~Belin, F--31400 Toulouse (France) }
\address[lab4]{ LAPP, IN2P3-CNRS, BP 110, F--74941 Annecy le Vieux (France) }

\thanks[label2]{M.D. is supported by a EU CMBNet fellowship.}

\begin{abstract}
We review the cosmological constraints put by the current CMB
experiment including the recent ARCHEOPS data, in the framework of
$\Lambda$-CDM and quintessence paradigm. We show that well chosen
combinations of constraints from different cosmological observations
lead to precise measurements of cosmological parameters. The Universe
seems flat with a 70 percents contribution of dark energy with an
equation of state very close to those of the vacuum.
\end{abstract}

\begin{keyword}
cosmology \sep Cosmic microwave background \sep cosmological parameters
\PACS 
\end{keyword}
\end{frontmatter}

\section{Introduction}
\label{intro}

The determination of cosmological parameters has always been a central
question in cosmology. In this respect the measurements of the
Cosmological Microwave Background (CMB) anisotropies on degree angular
scales has brought one of the most spectacular results in the field:
the flatness of the spatial geometry of the Universe, implying that
its density is close to the critical density.  Although, during the
last twenty years the evidence for the existence of non-baryonic dark
matter has strongly gained in robustness, observations clearly favour a
relatively low matter content somewhere between $20$ and $50\%$ of the
critical density, indicating that the dominant form of the density of
the universe is an unclustered form. Furthermore, the observations of
distant supernovae, at cosmological distance, provide a direct
evidence for an accelerating universe, which can possibly be explained
by the gravitational domination of a component with a relatively large
negative pressure: $P_\QUINT = w_\QUINT \rho_\QUINT$ with $w_\QUINT <
- 1 / 3$.  The cosmological constant $\Lambda$ (for which $w_\Lambda =
- 1$) is historically the first possibility which has been introduced
and which satisfies this requirement.  However, the presence of a
non-zero cosmological constant is a huge problem in physics due to the
``coincidence problem''.
In this paper we shortly summarise the different sets of data and
methods used to constrain cosmological parameters. We then conclude by
showing the results on cosmological parameters in both $\Lambda$-CDM
and quintessence paradigms. Such results are presented in details
in Beno{\^\i}t et al. 03b and Douspis et al. 03a \cite{us}.

\section{Method and data}
\label{data}

In the following, we make use of the most recent data available on the
CMB as well as on other relevant cosmological quantities in order to
examine constraints that can be set on cosmological parameters. We
assume Gaussian adiabatic fluctuations and a vanishing amount of
gravitational waves. Identically, a possible hot dark matter component
is neglected in the following. We investigate two cosmological
parameters sets: $\theta_1=(\Omega_{tot}, \Omega_\Lambda, \Obhh, h, n,
Q,\tau)$ for the $\Lambda$-CDM framework and $\theta_2=
(\OQ,w_\QUINT, \Obhh, h, n, \sigma_8)$ for the quintessence
paradigm. For the latter, we assume a flat Universe with no
reionisation and that $w_\QUINT = \CONST$ throughout all the epochs of
interest.  In order to use CMB data, we first reconstruct the
likelihood function of the various experiments.We follow the technique
developed in \cite{io} and used in \cite{us,io}, by constructing a
large $C_\ell$ power spectra database (CAMB: \cite{camb}). We
proceed by estimating cosmological parameters from the likelihood
functions reconstructed as described in \cite{us}.
We compute the value of the likelihood considering the actual band
powers dataset of COBE, BOOMERANG, DASI, MAXIMA, VSA, CBI, Archeops
(\cite{datacmb}) on each model of our grid.  For the different
combinations, we consider HST determination of the Hubble constant
 and supernov{\ae} determination of $\Omega_{m}$ and
$\Lambda$ (\cite{hstsn}). Then, for the quintessence paradigm
case we used the estimations of $\sigma_8$ leading to low values
(\cite{sigma8}), by considering: $\sig \OmM^{0.38} = 0.43 \pm 10\%$
($68\%$ C.L) because the cluster normalisation of the spectrum is highly
sensitive to the quintessence scenario.

\section{Cosmological Parameter constraints in $\Lambda$-CDM paradigm}        
\subsection{Archeops}\label{archonly}  

\begin{figure}[!ht]
\resizebox{!}{!}{\includegraphics[angle=0,totalheight=7cm,
        width=7cm]{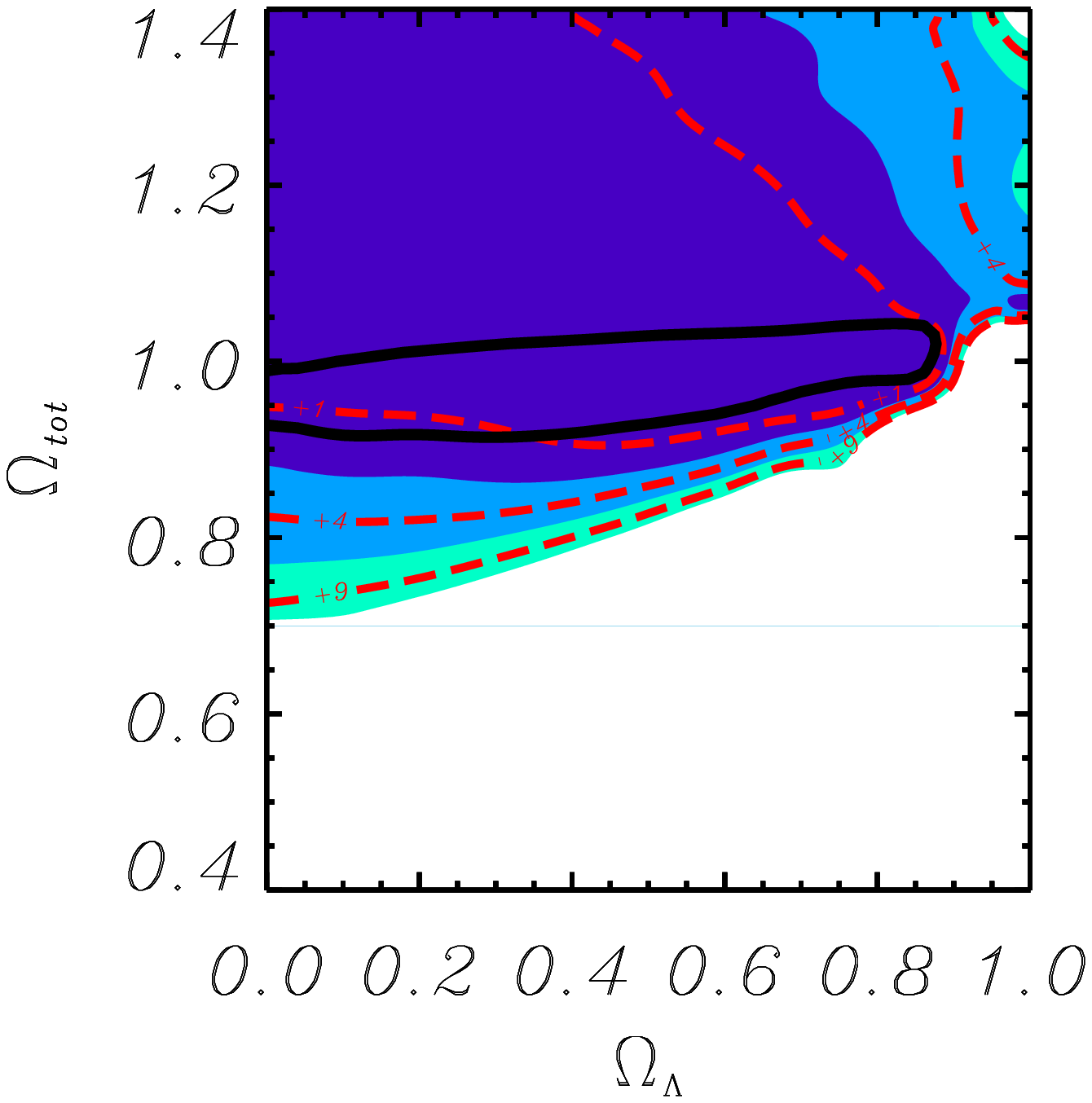}\includegraphics[angle=0,totalheight=7cm,
        width=7cm]{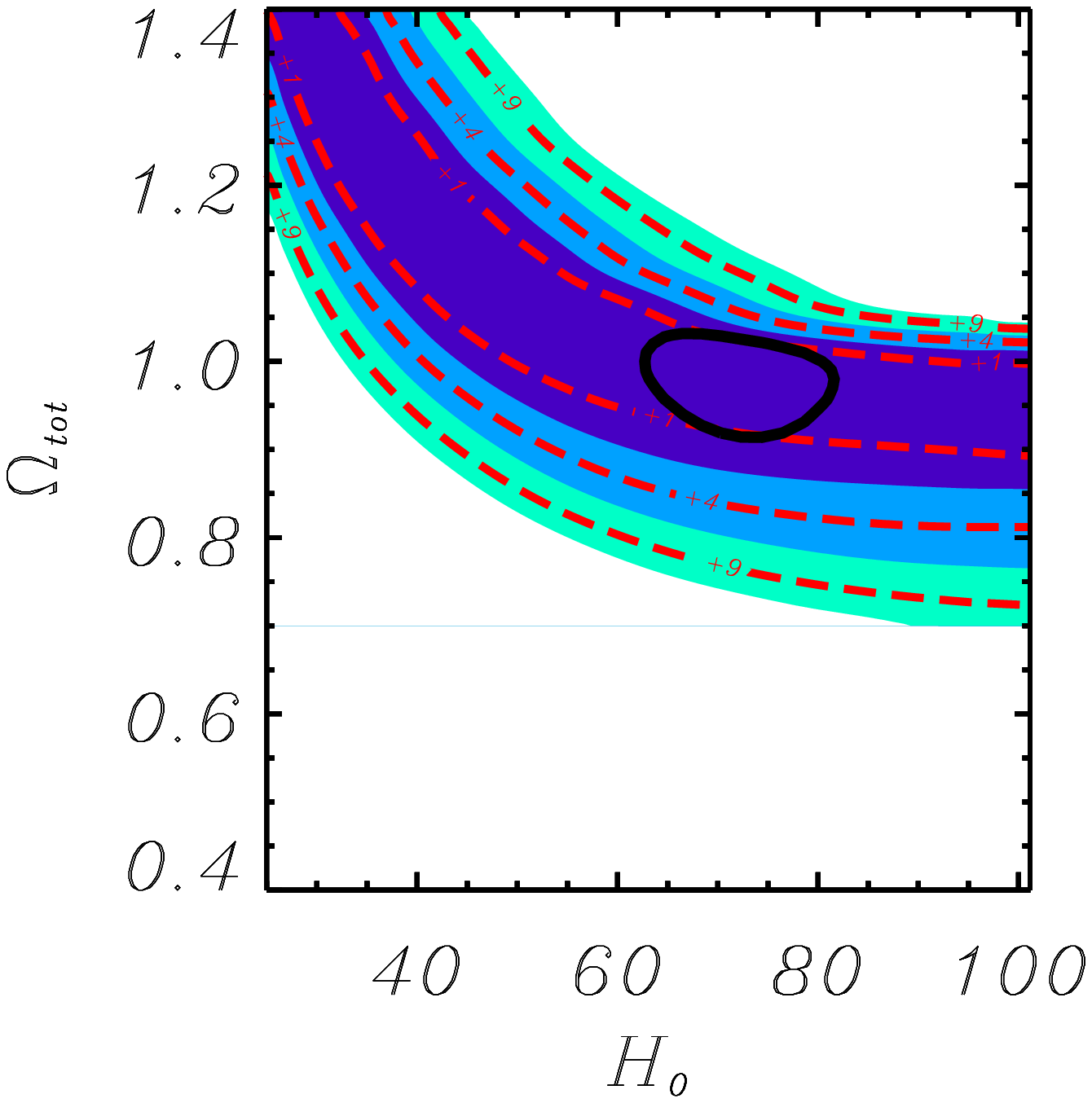}}

\caption{\label{arcmix}Likelihood contours  in the $(\OmL, \OmT)$
  (left) and $(H_0, \OmT)$ (right) planes using the Archeops dataset;
  the three coloured regions (three contour lines) correspond to resp.
  68, 95 and 99\% confidence levels for 2-parameters (1-parameter)
  estimates. Black solid line is given by the combination Archeops +
  HST, see text.}

\end{figure}

We first find constraints on the cosmological parameters using the
Archeops data alone. The cosmological model that presents the best fit
to the data has a $\chi^2_{gen}= 6/9$. Figure~\ref{arcmix} gives
confidence intervals on different pairs of parameters.  The Archeops
data constrain the total mass and energy density of the Universe
($\Omega_{tot}$) to be greater than 0.90, but it does not provide
strong limits on closed Universe models.  Fig.~\ref{arcmix} also shows
that $\Omega_{tot}$ and $h$ are highly correlated.
Adding the HST constraint for the Hubble constant leads to the tight
constraint $\Omega_{tot}= 0.96^{+0.09}_{-0.04}$ (full line in
Fig.~\ref{arcmix}), indicating that the Universe is flat.\\
Using Archeops data alone we can set significant constraints neither
on the spectral index $n$ nor on the baryon content $\Obhh$ because of
lack of information on fluctuations at small angular scales.

\subsection{Archeops and other CMB experiments}    

\begin{figure}[!t]
\resizebox{!}{!}{\includegraphics[angle=0,totalheight=7cm,
        width=14cm]{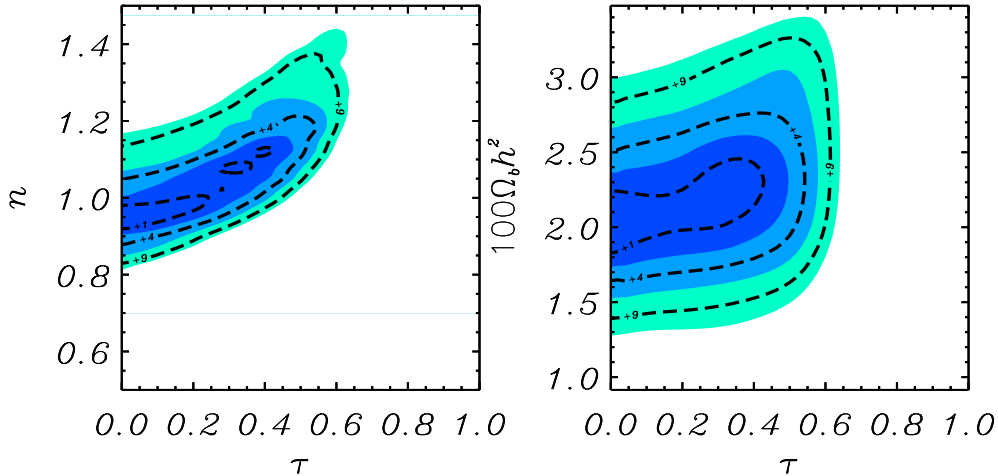}}
\caption{\label{tau}Likelihood contours in the $(\tau, n)$ and $(\tau,
\Obhh)$ planes using Archeops + CBDMVC datasets.}
\end{figure}

By adding the experiments listed in section~\ref{data} we now provide
the estimate of the cosmological parameters using CMB data only.  The
constraints are shown on Fig.~\ref{tau} and
\ref{alllow}~(left).  The combination of all CMB experiments provides
$\sim 10\%$ errors on the total density, the spectral index and the
baryon content respectively: $\OmT=1.15^{+0.12}_{-0.17}$,
$n=1.04^{+0.10}_{-0.12}$ and $\Obhh=0.022^{+0.003}_{-0.004}$.  These
results are in good agreement with recent analyses performed by other teams.
As shown in Fig.~\ref{tau} the spectral index and the optical depth
are degenerate. Fixing the latter to be $\tau < 0.20$, leads to
stronger constraints on both $n$ and $\Obhh$.  With this constraint,
the preferred value of $n$ becomes slightly lower than 1,
$n=0.96^{+0.03}_{-0.04}$, and the constraint on $\Obhh$ from CMB alone
is not only in perfect agreement with BBN determination but also has
similar error bars, $\Obhh_{(CMB)}=0.021^{+0.002}_{-0.003}$.  It is
important to note that many inflationary models (and most of the
simplest of them) predict a value for $n$ that is slightly less than
unity (see, e.g., \cite{lyth} for a recent review).

\subsection{Adding non--CMB priors}

\begin{figure}[!h]
\resizebox{!}{!}{\includegraphics[angle=0,totalheight=7cm,
        width=7cm]{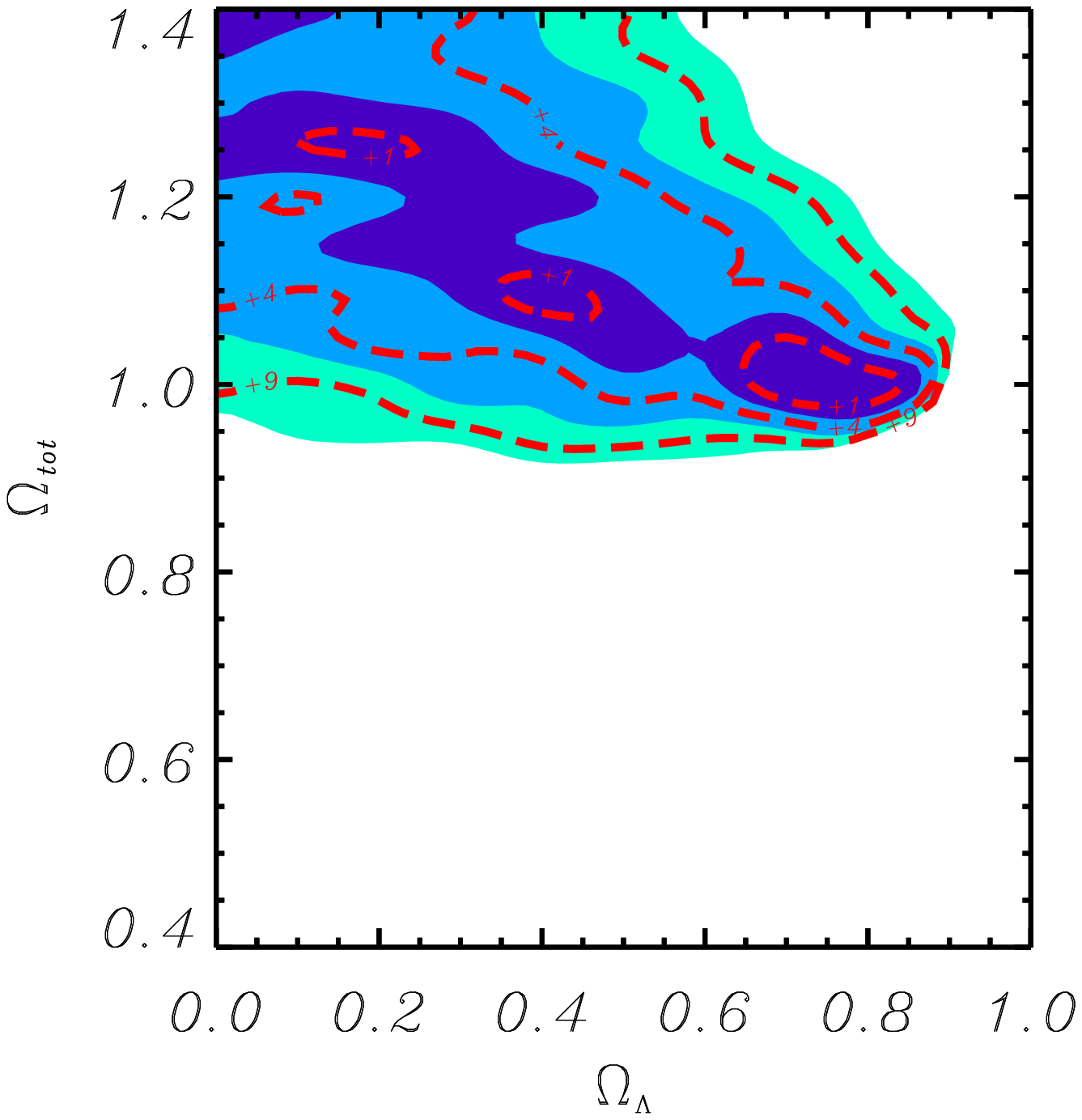}\includegraphics[angle=0,totalheight=7cm,
        width=7cm]{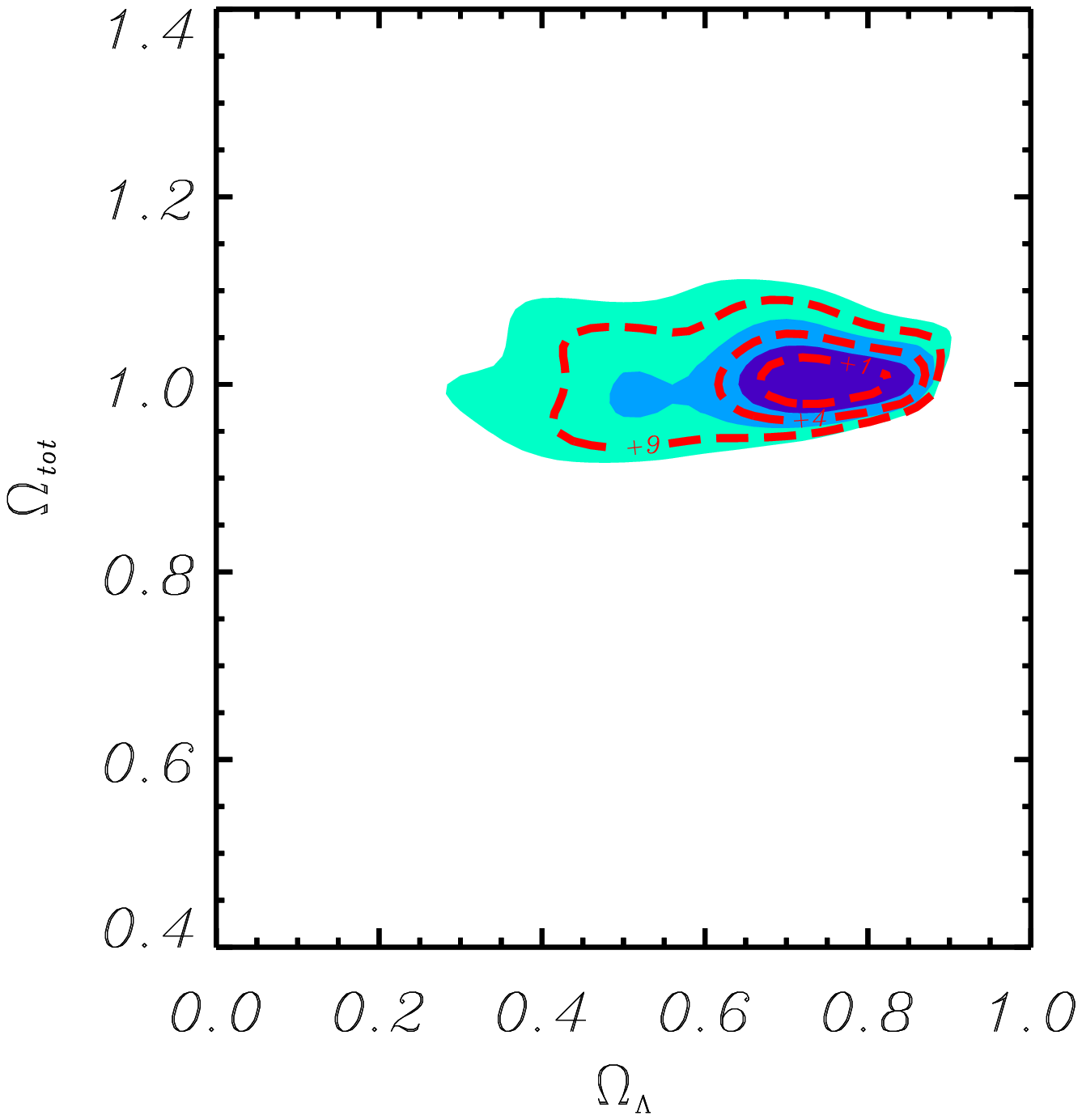}}
\caption{\label{alllow}Likelihood contours in the $(\OmT, \OmL)$. Left: constraints using Archeops+CBDMVC
 datasets. Right: adding HST prior for $H_0$.}
\end{figure}

In order to break some degeneracies in the determination of
cosmological parameters with CMB data alone, priors coming from other
cosmological observations are now added.  The results with the HST
prior are shown in Figure~\ref{alllow}~(right). Considering the
combination Archeops~+~CBDMVC~+~HST, the best model is $(\Omega_{tot},
\Omega_\Lambda, \Obhh, h, n, Q,
\tau)=(1.00,0.7,0.02,0.665,0.945,19.2{\rm \mu K},0.)$ with a
$\chi^2_{gen}=41/68$.  The constraints on $h$ break the degeneracy
between the total matter content of the Universe and the amount of
dark energy as discussed in Sect.~\ref{archonly}.  The constraints are
then tighter as shown in Fig.~\ref{alllow}~(right), leading to a value
of $\Omega_\Lambda = 0.73^{+0.09}_{-0.07}$ for the dark energy content, in
agreement with supernov{\ae} measurements if a flat Universe is
assumed.

\newpage

%

%

\vspace*{0.5cm}

\section{Cosmological Parameter constraints in Quintessence paradigm}

\subsection{CMB alone}

\begin{figure}[!h]
\begin{center}
\resizebox{!}{!}{\includegraphics[angle=0,totalheight=7cm,
        width=7cm]{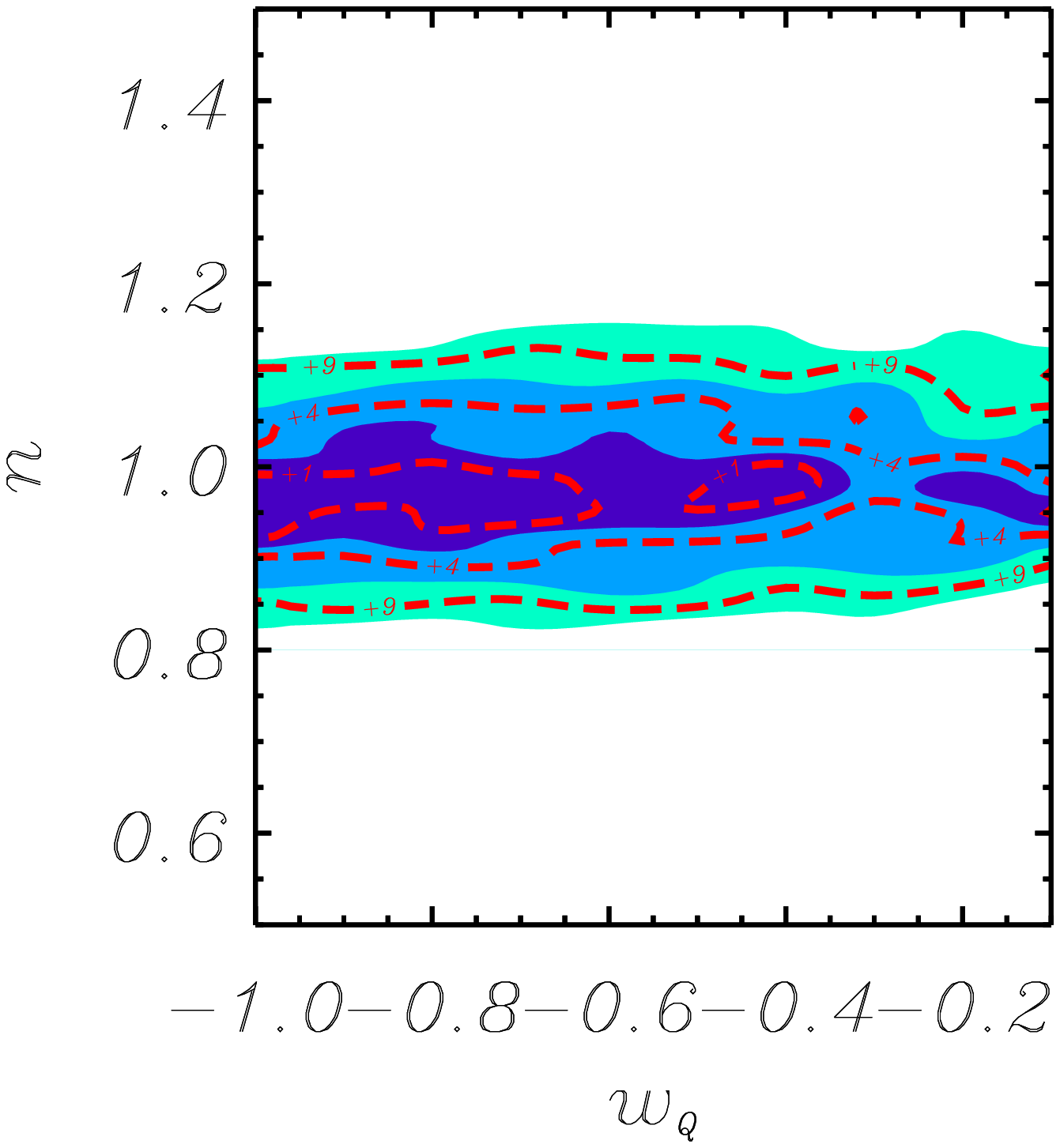}\includegraphics[angle=0,totalheight=7cm,
        width=7cm]{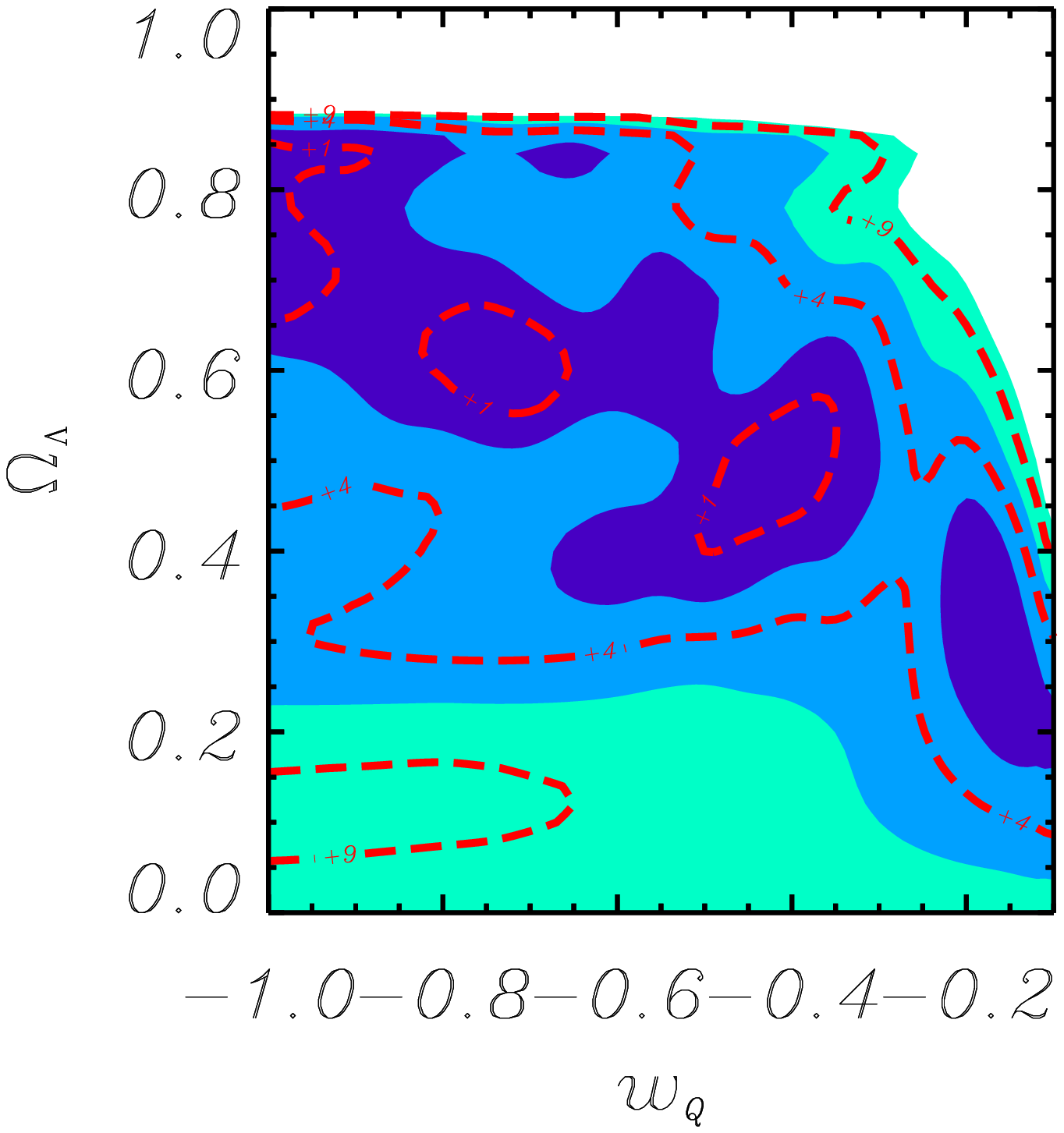}}
\end{center}
\caption{Present CMB dataset likelihood contours in the quintessence
paradigm. The sharpness of contours at $\OQ = 0.9$ is due to grid
effect.}
\label{cmbalone}
\end{figure}

Constraints given by the CMB on some of our investigated parameters
are shown in Fig.~\ref{cmbalone}. Considering only CMB constraints
leads to degeneracies between parameters. Fig.~\ref{cmbalone} shows
the case of one parameters, $n$, which is not affected by
the assumed equation of state of the dark energy ($\Omega_\BAR$ is not either). The preferred
value and error bars are $n = 0.95 \pm 0.05$ ($68\%$ C.L.) and
$\Omega_\BAR h^2 = 0.021 \pm 0.003$.  Using CMB alone leaves the
2-parameters space ($\OQ, w_\QUINT$) almost unconstrained.   In our
analysis, we found that with the improvement of CMB data obtained by
the addition of Archeops band powers reduces appreciably the contours
of constraints on the quintessence parameters as well as on
cosmological parameters because of the position and the amplitude of
the first acoustic peak are better determined, but still does not
allow to break the degeneracies.

\subsection{Adding Non CMB priors}

\begin{figure}[!h] 
\begin{center}
\resizebox{!}{!}{\includegraphics[angle=0,totalheight=7cm,
        width=7cm]{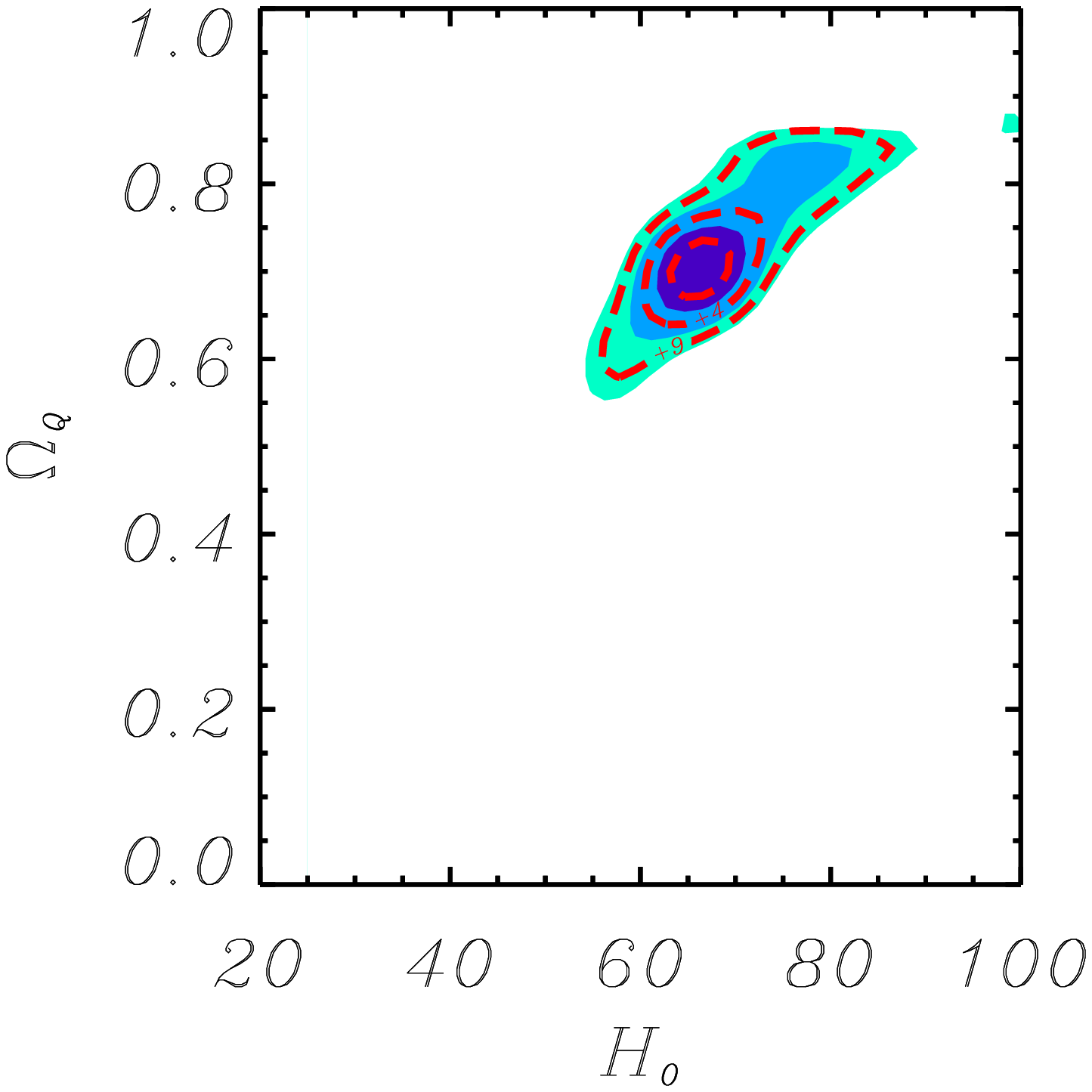}\includegraphics[angle=0,totalheight=7cm,
        width=7cm]{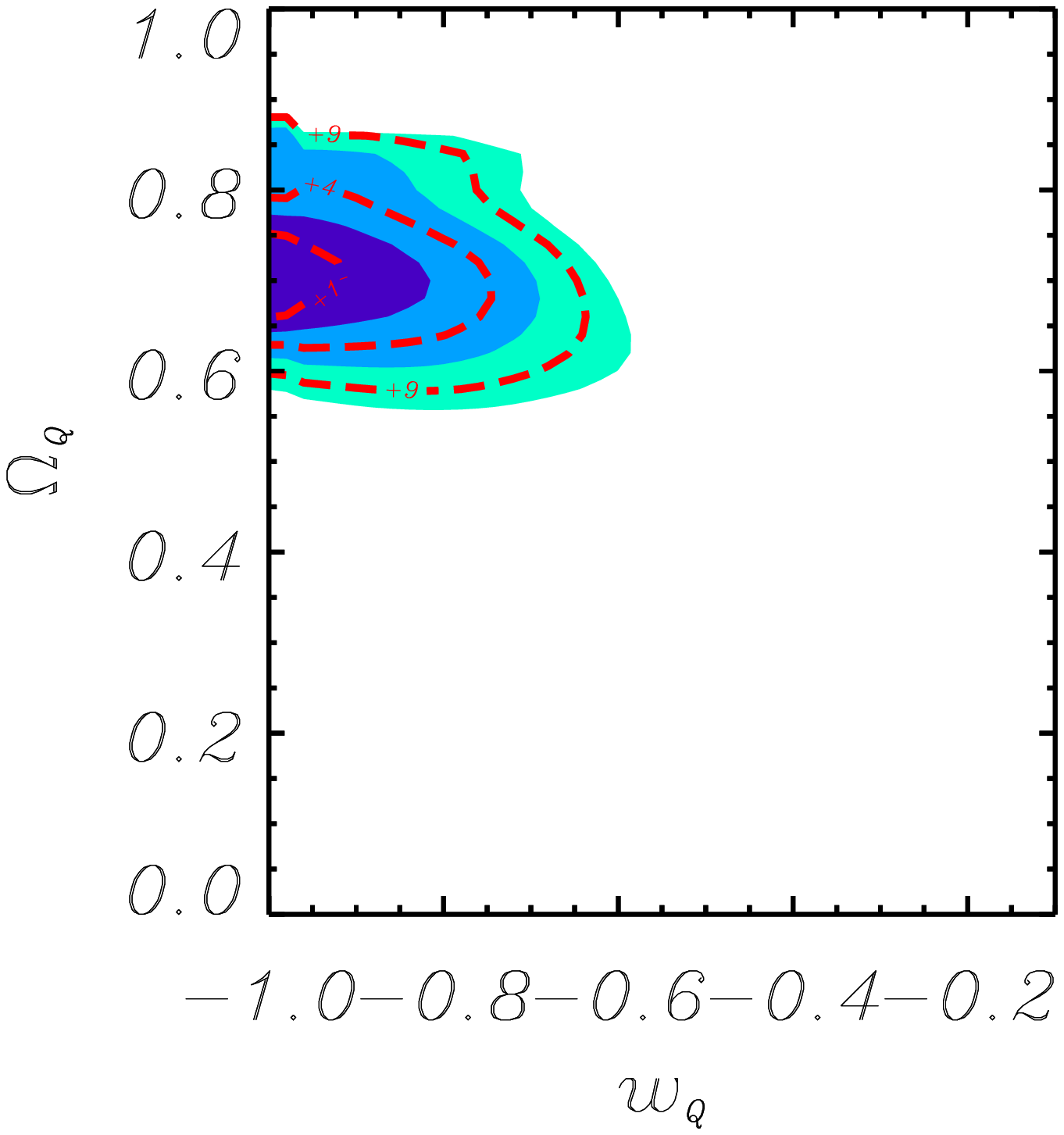}}
\end{center}
\caption{\label{fin}Likelihood contours with CMB + all priors in the quintessence paradigm.}

\end{figure}

In order to break the degeneracy it is clearly necessary to consider
the additional information on the normalisation of the spectrum by the
value of $\sig$, which is highly sensitive to the equation of state
(see section 2 of Douspis et al. 03a).  We can furthermore consider the
angular distance coming from distant supernovae. Assuming a flat
cosmology, the information on the luminosity of the supernovae can be
expressed in term of constraints on the dark energy density and
equation of state.  Finally the Hubble constant determination by HST
Key project is also considered.
Combining all the priors finally allows to put
strong constraints on both quintessence parameters (Fig.~\ref{fin}): $\OQ = 0.70^{+0.10}_{-0.17}$, $w_\QUINT =
-1^{+0.25}$ ($95\%$ C.L.) and finally breaks the $(H_0, \OQ)$
degeneracy.\\
As a main result, it appears that the classical $\Lambda$-CDM is then
comforted and given the priors we used there is no need for
quintessence to reproduce the present data.

\section{Conclusion}

We have studied the constraints that can be obtained on cosmological
parameters within the $\Lambda$-CDM and the quintessence paradigms by
using various combination of observational data set.\\
Our analysis method has been to investigate
contours in 2D parameters space. Such an approach allows to examine
possible degeneracies among parameters which are not easy to identify
when constraints are formulated in term of single parameter. For
instance we found that CMB data alone, despite the high precision data
obtained by Archeops do not require the existence of a non-zero
contribution of quintessence, because of the degeneracy with the
Hubble constant: in practise CMB data leave a large fraction of the
$\Omega_Q$--$w_\QUINT$ plane unconstrained, while only a restricted
region of the $\Omega_Q$--$H_0$ is possible. On the contrary we found
that almost no correlation exist with the baryonic content
$\Omega_\BAR$ nor the primordial index $n$. In order to restrict the
parameter space of allowed models we have applied several different
constraints. Interestingly, we found that the amplitude of the dark
matter fluctuations, as measured by clusters abundance or large scale
weak lensing data can potentially help to break existing degeneracies,
although existing uncertainties, mainly systematics in nature do not
allow firm conclusion yet. Clearly this will be an important check of
consistency in the future. We have then added constraints from
Supernovae data as well as HST estimation of the Hubble constant in
order to break existing degeneracies. This allows us to infer very
tight constraints on the possible range of equation of state of the
dark energy. Probably the most remarkable result is that no preference
for quintessence  does emerge from existing CMB data.



\begin{thebibliography}{00}
\bibitem{us} Beno{\^ \i}t, A.~et al.\ 2003,  A\&A, 399, L25 (Archeops collaboration), Douspis M. \ETAL{} 2003a , A\&A in press, astroph/0212097


\bibitem{io} Bartlett, J.\ G., Douspis, M., Blanchard, A., \& Le Dour, M.\ 2000, A\&AS, 146, 507, Douspis M., Bartlett J.G. \& Blanchard A., 2003, A\&A in press 



\bibitem{datacmb}Beno{\^ \i}t, A.~et 
al.\ 2003, A\&A, 399, L19,  Halverson, N.~W. \ETAL{} 2002, MNRAS, 568, 38, Lee, A.~T. \ETAL{} 2001, ApJ, 561, L1, Netterfield, C.~B. \ETAL{} 2002, ApJ,
571, 604,  Pearson, T.~J. \ETAL{}  
2002, astro-ph/0205388 , Scott, P. F.\ETAL{} 2002,  astro-ph/0205380, Tegmark, M.~1996, ApJ, 464, L35

\bibitem{hstsn}Freedman, W.~L., Madore, B.~F., 
 Gibson, B.~K.,~{et~al.}~2001, ApJ, 553, 47, Perlmutter, S., Aldering, G., Goldhaber, G., {et~al.}~1999, ApJ, 517,
565 \& {\tt http://www-supernova.lbl.gov}

\bibitem{camb}
Lewis, A., Challinor, A., \& Lasenby, A.~2000, ApJ, 538, 473

\bibitem{sigma8}  Reiprich, 
T.~H.~\& B{\" o}hringer, H.\ 2002, ApJ, 567, 716 , Seljak, U,
astro--ph/0111362, Viana, P.~T.~P., Nichol, R.~C., \& Liddle, A.~R.\
2002, ApJ letters, 569, L75


\bibitem{lyth} Lyth, D. H. \& Riotto, A.~1999, Phys.Rep., 314, 1






\end{thebibliography}
\end{document}